\documentclass[prl,aps,twocolumn,showpacs,floatfix]{revtex4}
\usepackage{graphicx}
\usepackage{bm}
\usepackage{amsmath}
\usepackage{amssymb}
\usepackage{amsfonts}
\usepackage{dcolumn}%
\usepackage{bm}
\usepackage{color}

\newcommand{\ad}[1]{\tilde{#1}}

\begin{document}

\title{Mechanical fluctuations suppress the threshold of soft-glassy solids : the secular drift scenario}

\author{Adeline Pons$^{1}$}
\author{Axelle Amon$^{2}$\footnote{axelle.amon@univ-rennes1.fr}}
\author{Thierry Darnige$^{1}$}
\author{J\'er\^ome Crassous$^{2}$}
\author{Eric Cl\'ement$^{1}$}

\affiliation{$^{1}$PMMH, ESPCI, UMR CNRS 7636 and Universit\'e Paris 6 \& Paris 7, 75005 Paris, France}
\affiliation{$^{2}$Universit\'e Rennes 1, Institut de Physique de Rennes (UMR UR1-CNRS 6251), B\^{a}t.~11A,Campus de Beaulieu, F-35042 Rennes, France}

\date{\today}
\begin{abstract}
We propose a dynamical mechanism leading to the fluidization by
external mechanical fluctuations of soft-glassy amorphous material
driven below the yield-stress. The model is based on the combination
of memory effect and non-linearity, leading to an accumulation of tiny
effects over a long-term. We test this scenario on a granular packing
driven mechanically below the Coulomb threshold. We bring evidences
for an effective viscous response directly related to small stress
modulations in agreement with the theoretical prediction of a generic
secular drift. We finally propose to extend this result more
generally, to a large class of glassy systems.
\end{abstract}
\date{\today}
\maketitle

Numerous amorphous materials such as concentrated suspensions,
colloidal glasses, foams or granular materials share common global
features in their mechanical response to shear
\cite{Liu98,Berthier2001}. They are characterized by a yield stress
below which the material appears as a
solid~\cite{Coussot2002,Moller2009}.  As this behaviour is shared by
so many different materials, several conceptual and theoretical
frameworks emerged
recently~\cite{Sollich1997,Falk1998,Hebraud1998,Fielding2000,Derec2001,Bocquet2009}
to provide a quantitative basis for the phenomenology of soft glassy
rheology (SGR) above and beyond the yield stress. Even though many
parallel approaches exist, sometimes at different level of
description, they all share either explicitly or implicitly, the
underlying idea that mesoscopic collective processes triggered by
thermal or mechanical activation, contribute to the material
fluidity. The direct visualization of local plastic events and the
associated complex avalanching dynamics is supported by many
experimental \cite{Kabla2007,Schall2007,Amon2012,Lebouil2014} or
numerical~\cite{Maloney2006,Tanguy2006,Chaudhuri2013} studies. In the
``solid phase`` corresponding to a strong dynamical arrest,
soft-glassy systems display ageing properties manifesting in a slow
creep relaxation
process~\cite{Cloitre2000,Derec2003,Nguyen2011,Siebenburger2012}. Ageing
properties stem from a remaining thermal activation providing the
possibility to cross enthalpic or entropic barriers and progressively
set the system into deeper local minima where mechanical solidity is
reinforced. The existence of external mechanical noise was also
proposed as a substitute for thermal activation. In this sense, the
behaviour of these amorphous soft glassy solids is very close
phenomenologically to molecular glass-formers obtained by thermal
quenching~\cite{Debenedetti2001}. Yet, the fact that such mechanical
noise truly acts as an effective temperature is presently
debated~\cite{Nicolas2014} and indeed, deep differences in the way
thermal noise and mechanical fluctuations act in amorphous systems has
been recently pointed out~\cite{Agoritsas2015}.\\
In the solid glassy phase, where the system never reaches thermal
equilibrium at the level of experimental times, in addition to the
presence of an elastic response, theories have to account for the loss
of ergodicity. This is done either by introducing memory kernel in the
soft-glassy rheology \cite{Fielding2000,Siebenburger2012,
  Voigtmann2013} or by providing phenomenologically, new dynamical
relations for an effective fluidity parameter ~\cite{Derec2001,
  Derec2003} suited to render the ''rheological age'' of the system
and its temporal evolution. Note that the two approaches, not working
at the same level of representation, are not necessarily contradictory
and in some simple cases, explicit connections can even be made
\cite{Derec1999}.\\
On a practical point of view, even far from the fluidization
thresholds, a lot of situations show that vanishingly small
perturbations cannot be neglected in presence of a bias. Since such
effects may be cumulated over long times, it becomes problematic when
a solid response is expected but uncontrolled mechanical noise would
eventually lead to a significant creep. For example, the effect of
mechanical noise on soils is of major importance for the long-term
stability of structure foundations ~\cite{Murayama1984}. It may also
play a determinant role in the context of earthquakes
triggering~\cite{Johnson2005}.  Controlled mechanical fluctuations can
also be used as an investigation tool, as for example in superposition
rheology~\cite{Negi2010}. In this instance, understanding the system
response to various forcing of different forms and amplitudes and
also, the importance of inherent apparatus wobbling noise, is crucial
for an accurate exploitation of the system dynamics. On a theoretical
point of view, it has been shown that ageing in a glass spin model is
interrupted in presence of a bias~\cite{Berthier2000}.\\
In this letter, we propose a new conceptual picture to understand a
fluidization process that a soft glassy materials may undergo in the
solid phase, under external mechanical noise. This scenario differs
from an activated process and does not require the introduction of an
effective temperature. First, theoretical arguments are presented to
describe the solid phase where ageing and shear rejuvenation processes
are both present. Second, an explicit derivation is presented on a
generic rheological model. Third, we present experiments on granular
packing sheared below the Coulomb threshold and we show that the
response to small mechanical modulations is in agreement with the
generic predictions of the model.  Finally, the result's generality
and it application to soft glassy materials is discussed.\\
%
--{\it Model ingredients.} Models aiming at describing yield-stress
fluids and amorphous materials in the solid phase, need to account for
two fundamental features in their
dynamics~\cite{Sollich1997,Derec2001}: (i) ageing of the system with
time and (ii) rejuvenation due to shear rate $\dot{\gamma}$. That
rejuvenation can be seen microscopically as a local structural
reorganizations induced by the strain. When an amorphous material is
submitted to a cyclic load, after a complete cycle, the system is not
back to its initial state~\cite{Derec2001}, which means that the rate
of evolution of the variable describing the state of the system has an
even dependence on the shear rate $\dot{\gamma}$.  Because of the
different time scales at play in the dynamics (typical time of
reorganization compared to the ageing time), the description of glassy
materials depends on the observation time scale. Choosing this scale
can be problematic in creep experiments as the system exhibits no
intrinsic time scale. In the presence of stress fluctuations, the
macroscopic variables measured are averaged quantities giving the mean
long-term behaviour. If the system is submitted to stress variations
of typical amplitude $\delta$, very small when compared to the yield
stress $\sigma_D$, and displaying a characteristic time $\tau_{vib}$,
a pertinent observation time is given by the number of perturbations
of amplitude $\delta$ necessary to accumulate an equivalent stress of
order $\sigma_D$: $T_{obs} = \frac{\sigma_D}{\delta}
\tau_{vib}$. Because of the positive non-linear dependence of the
rejuvenation term, one can expect a dynamical stack-up of those
perturbations, giving rise after a time of order $T_{obs}$, to an
equivalent stress of order $\sigma_D$.

--{\it Fluidization as a secular drift.} In order to demonstrate
simply how this mechanism works, we build on the macroscopic
rheological model proposed by Derec {\it et al.}~\cite{Derec2001} to
understand the rheology of soft glassy materials. This model was used
to analyse ageing and non-linear rheology of pastes \cite {Derec2003}
and also creeping processes in granular matter \cite {Nguyen2011,
  Amon2012, Espindola2012}. This generic model introduces a
macroscopic phenomenological variable, the \emph{fluidity} defined as
the inverse time scale characterizing the material visco-elastic
response. To provide a comprehensive analytical understanding of how a
steady fluidity can appear below the dynamical yield stress
$\sigma_D$, we first study the response on the simplest non trivial
form of the model :
\begin{eqnarray}
\overset{.}\sigma &=&G\overset{.}{\gamma
}-f\sigma \label{EquFluStress}\\
\dot{f}&=&-af^2+r\dot{\gamma}^2, \label{eqfluidity2}
\end{eqnarray}
with $\sigma$ the applied shear stress, $\dot{\gamma}$ the shear rate
and $G$ the shear elastic modulus, $f(t)$, is the
fluidity. Dimensionless parameters $a$ and $r$ represent respectively
ageing and shear-induced rejuvenation processes and for clarity and
simplicity, we consider them as constant. The stationary solutions of
those equations depend on the value of the stress compared to the
dynamical yield threshold, $\sigma_D = G \sqrt{\frac{a}{r}}$. For a
constant $\sigma < \sigma_D$, the fluidity $f$, as well as the shear
rate $\dot{\gamma}$, decreases to 0 as the inverse of time, thus
leading to a logarithmic creep process. We consider the case of a mean
imposed stress $\sigma_0$ below the threshold $\sigma_D$ combined with
a modulation of small amplitude $\delta \ll \sigma_0$, leading to an
imposed stress $\sigma(t) = \sigma_0 + \delta \sin (\omega t)$. By
construction, the present fluidity model has no time scale. When
imposing a modulation, one can study the in-phase and out-of-phase
responses~\cite{Derec2001} over a time of the order of $\tau_{vib} =
\frac{2 \pi}{\omega}$. Here we aim at understanding the long-term
behaviour, given by the number of cycles of amplitude $\delta$
necessary to build-up an equivalent stress of order $\sigma_D$:
$T_{obs} = \tau_{vib} / \epsilon$, with $\epsilon = \delta /
\sigma_D$. The equations are adimensionalized using the following
scales: $1/(\epsilon \omega)$ for time, $\sigma_D$ for stress and
$\gamma_0=\sigma_D/G$ for deformation. The adimensionalized variables
are written with a tilde thus yielding the equations:
$\dot{\tilde{\sigma}} =
\dot{\tilde{\gamma}}-\tilde{f}\tilde{\sigma} \label{eq.fluidity1bis}$
and $\dot{\tilde{f}} = -a( \tilde{f}^2 -
\dot{\tilde{\gamma}}^2) \label{eq.fluidity2bis}$, with $\tilde{\sigma}
= \tilde{\sigma_0} + \epsilon \sin \left( \frac{\tilde{t}}{\epsilon}
\right)$, which gives :
\begin{equation}
\dot{\tilde{\gamma}} = \cos \left( \frac{\tilde{t}}{\epsilon} \right)
+ \tilde{f} \tilde{\sigma_0} + \epsilon \tilde{f} \sin \left(
\frac{\tilde{t}}{\epsilon} \right)
\end{equation}
Dynamically, one obtains a two-times system with $T = \tilde{t}$, the
time of observation corresponding to creep (\emph{slow} time) and the
modulation time $\tau = \tilde{t}/\epsilon$ (\emph{fast} time). A
multiple scale perturbation analysis can be done~\cite{Hinch1991},
using $\frac{d}{d\tilde{t}} = \frac{1}{\epsilon}
\frac{\partial}{\partial \tau} + \frac{\partial}{\partial T}$ and
searching a solution of the form $\tilde{f}(\tau,T) =
\tilde{f}^{(0)}(\tau,T) + \epsilon \tilde{f}^{(1)}(\tau,T)+ \dots$. We
then obtain for $(1 - \tilde{\sigma}_0^2) = O(1)$:
\begin{eqnarray}
& & \frac{1}{\epsilon} \frac{\partial \ad{f}^{(0)}}{\partial \tau} +
  \frac{\partial \ad{f}^{(1)}}{\partial \tau} + \frac{\partial
    \ad{f}^{(0)}}{\partial T} + O(\epsilon) = -a \left[- \frac{1 -
      \cos 2 \tau}{2} \right. \nonumber \\ & & \left. + (1 -
    \ad{\sigma}_0^2) \left(\ad{f}^{(0)}\right)^2 + 2 \ad{f}^{(0)}
    \ad{\sigma}_0 \sin \tau + O(\epsilon) \right]
\end{eqnarray}

\begin{figure}[h]
\includegraphics[width=.65\linewidth]{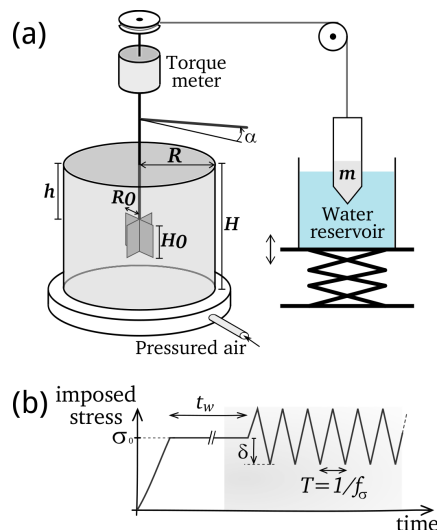}
\caption{$(a)$ Experimental set-up. $(b)$ Imposed stress during an
  experiment: stress ramp to reach mean stress, $\sigma_0$, constant
  stress during $t_w$ and stress modulation characterized by a
  frequency $f_\sigma$ and an amplitude $\delta$.}
\label{fig_setup}
\end{figure}

\begin{figure}[h]
\includegraphics[width=.75\linewidth]{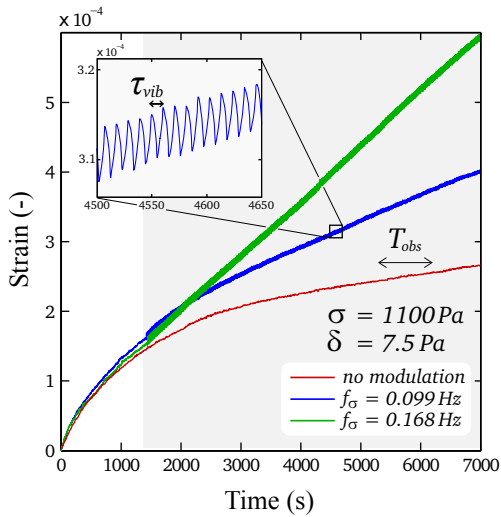}
\caption{Strain as a function of time for three experiments performed
  at $\sigma_0 = 1100$~Pa, and $\delta = 7.5$~Pa, and for various
  oscillations frequencies.  The oscillations start at t = 1500s (grey
  area).}
\label{fig_creep}
\end{figure}

\begin{figure}[h]
\includegraphics[width=.75\linewidth]{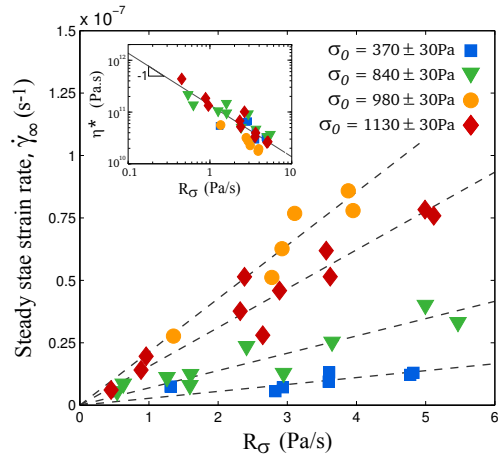}
\caption{Main plot: Steady state strain rate, $\dot{\gamma}_\infty $,
  as a function of modulation stress rate $R_\sigma$ for four
  different values of mean strees, $\sigma_0$. The straight lines are
  linear fits. Inset: Normalized viscosity $\eta^* = \eta/\sqrt{1
      - \frac{\sigma_0^2}{\sigma_D^2}}$ as a function of $R_\sigma$
    with $\sigma_D = 1240$ Pa.}
\label{Gp}
\end{figure}

From the leading order $O(\frac{1}{\epsilon})$, one obtains
$\frac{\partial \ad{f}^{(0)}}{\partial \tau} = 0$, so that
$\tilde{f}^{(0)}(\tau,T) = \ad{f}^{(0)}(T)$: the envelope is a
function of the slow time only. The order $O(1)$ gives:
\begin{eqnarray}
\frac{\partial \ad{f}^{(1)}}{\partial \tau} &=& - \frac{d
  \ad{f}^{(0)}}{dT} + a \left[\frac{1}{2} - (1 - \ad{\sigma}_0^2)
  \left(\ad{f}^{(0)}\right)^2 \right] \nonumber \\ & & -a \left[ 2
  \ad{f}^{(0)} \ad{\sigma}_0 \sin \tau - \frac{\cos 2 \tau}{2}
  \right] \label{eq.O1}
\end{eqnarray}
The term $ - \frac{d \ad{f}^{(0)}}{dT} + a \left[\frac{1}{2} - (1 -
  \ad{\sigma}_0^2) \left(\ad{f}^{(0)}\right)^2 \right]$ in the rhs of
equation~\ref{eq.O1} does not depend on $\tau$ so that its integration
would give a term $\propto \tau$ which would lead to a failure of the
expansion on long time. This term, called the secular term because its
effect is seen only after a very long time, thus needs to be cancelled
for the perturbation analysis to hold (see
e.g.~\cite{Hinch1991}). This leads to the differential equation:
\begin{equation}
\frac{\partial \ad{f}^{(0)}}{\partial T} = a \left[\frac{1}{2} - (1 -
  \ad{\sigma}_0^2) \left(\ad{f}^{(0)}\right)^2 \right]
\end{equation}
This equation corresponds to the normal form of a saddle-node
bifurcation $\dot{x} = \mu - x^2$. For $\mu > 0$ the solution
$+\sqrt{\mu}$ is the only stable stationary solution. The dimensional
expression of the stationary value for the mean fluidity, $f^*$, is
thus: $f^* = \frac{\omega \frac{\delta}{\sigma_D}}{\sqrt{2 \left(1 -
    \frac{\sigma_0^2}{\sigma_D^2} \right)}}$, which is finite for
non-vanishing values of $\delta$, the stress modulation, even when
$\sigma_0 \ll \sigma_D$. Consequently, even far below the yield
threshold, the long-term behaviour tends to create a liquid-like
response, with constant mean strain rate $\dot{\gamma}^{(0)} =
\frac{f^*}{G} \sigma_0$ corresponding to a finite effective viscosity:
\begin{equation}
\eta = G/f^*= G\sqrt{2 \left(1 -
    \frac{\sigma_0^2} {\sigma_D^2} \right)}\frac{\sigma_D}{\omega \delta}
\label{eq:visc}
\end{equation}
Note that this viscous response is linearly related to the inverse of
the stress modulation rate ($R_{\sigma}=2\omega \delta/\pi$). In
Appendix we numerically show that secular drift is a robust result
that can be applied to a large class of macroscopic rheological
models, the essence of the phenomenon being indeed captured by the
previous simple case. We also show that the secular drift do not
depends on the nature of the stress modulations (see results for
random forcing in Appendix). \\
--{\it Stress modulation experiments.} The generic theoretical
outcomes are now tested experimentally on a granular packing under a
confinement pressure that sets a scale for the Coulomb dynamical yield
stress.  Granular materials are often seen as rigorously
athermal. Indeed, in most numerical approaches, granular contacts are
modelled as elastic repulsive forces and a Coulomb solid friction
threshold. Consequently, for infinitesimal deformations around a
reference state, a granular packing should possess a true elastic
response and displays no ageing~\cite{Dagois2012}. Note however, in
the limit of very small if not zero friction the establishment of a
linear elastic response under finite shear is questionable
\cite{Combe2000}. Moreover, for real granular materials, the actual
pressures at contact are generically high and contacts may creep
plastically. Therefore, the contact status will be intrinsically
coupled to a thermally activated process~\cite{Divoux2010}. In
addition, the contact status can also be extremely sensitive to the
ambient mechanical noise. In fact, real granular packing in the solid
phase, display ageing and shear rejuvenation that can be modelled
directly by equation \ref{eqfluidity2} \cite{Nguyen2011}.  Moreover,
the fluidity variable $f(t)$ was identified experimentally (for shear
stresses not too close to the yield stress), as the rate of occurrence
of local rearrangements called ''hot-spots''~\cite{Amon2012}, thus
providing an explicit experimental connection with more mesoscopic
theories describing structural relaxation processes.  As a
consequence, experiments on granular packing in the solid phase, can
be considered as of general relevance to the class of soft-glassy
materials that display similar phenomenology
\cite{Voigtmann2014}.\\
An experimental key point here is to achieve shear stress modulations
around a nominal value without introducing uncontrolled mechanical
perturbations. Besides residual external noise, always present, even
in quiet environments, a substantial source of mechanical noise comes
from motorized elements. This is why we designed the experimental
system as an Atwood machine. The set-up is shown on
Fig.~\ref{fig_setup}(a). It consists in a shear cell (Radius $R=5cm$,
height $H=10cm$) filled with glass beads of density
$\rho=2500\ kg/m^3$ and mean diameter $d=(200 \pm 30)\ \mu m$. A
well-defined packing fraction $\phi = 0.605\pm0.005$ is obtained by a
procedure described elsewhere~\cite{Nguyen2011}.
Shear is obtained by applying a torque on a four-blade vane ($R_0
=1.27\ cm$, $H_0 =2.54\ cm$) using a mass $m$ suspended from a pulley
(see fig.~\ref{fig_setup}(a), vane penetration $h=5\ cm$). A torque
probe measures the applied torque $\mathcal{T}$ and the angle of
rotation of the vane $\alpha$ is measured by an induction probe. We
defined the mean stress and the mean strain as
$\sigma=\frac{\mathcal{T}}{2\pi R_0^2H_0}$ and $\gamma=\frac{\alpha
  R_0}{R-R_0}$ respectively. In conditions of the present experiment,
the Coulomb threshold was determined at a value $\sigma_Y =~2300Pa
$. When a constant stress $\sigma_0$ smaller than the yield stress is
applied on the granular packing, a creep behaviour is observed with a
logarithmic dependence of the strain with time (red curve of
Fig.~\ref{fig_creep}). This behaviour was studied in~\cite{Nguyen2011}
and the fluidity model discussed in the previous part describes
accurately the observed response.
By variation of the Archimede's forces, a modulation of the applied
torque is obtained by vertical oscillatory displacements of a mass $m$
hanging partially in a water tank. The protocol
(fig.~\ref{fig_setup}(b)) is then: $i)$ stress ramp at constant stress
rate ($\dot{\sigma} =5 \ Pa/s$) up to the desired mean stress value
$\sigma_0$; $ii)$ constant shear $\sigma_0$ applied during
$t_w=1500\ s$; $iii)$ modulation of the stress around $\sigma_0$ for
at least 2 hours. The modulations are triangular oscillations of
amplitude $\delta$ and frequency $f_\sigma$.
Figure.~\ref{fig_creep} shows typical deformations for two experiments
performed at the same mean stress ($\sigma_0 = 1100 Pa$) and
oscillation amplitude ($\delta = 7.5Pa$) but for various oscillation
frequencies. During the constant stress phase, a slow increase of the
deformation, $\gamma(t)$, is observed corresponding to the beginning
of the logarithmic creep. Then, when submitted to oscillations, the
system will transit to a linear creep regime characterized by a constant
mean strain rate, $\dot{\gamma}_\infty $, which increases with the
oscillation frequency. The slope of this linear creep allows to define
an effective viscous response: $\eta = \sigma_0 /
\dot{\gamma}_\infty$.
Figure~\ref{Gp} shows the values obtained for the mean strain rate
$\dot{\gamma}_\infty$ as a function of the modulation stress rate
$R_\sigma = 2 \omega \delta/\pi$, for a given value of the applied
mean stress $\sigma_0$. The observed linear dependences are in
agreement with the normalization parameters chosen. Indeed, the finite
viscosity that we expect to arise from the modulation should vary as
$\eta \propto \frac{1}{\delta \omega}$ (see eq.~\eqref{eq:visc})
leading to a strain rate $\dot{\gamma}_\infty \propto \omega
\delta$. $\sigma_0$ corresponds to an applied shear stress far enough
from the dynamical threshold. Experimentally, when this limit is
approached one observes a strong increase of the strain rate. The
results are then much less reproducible and may be quite sensitive to
uncontrolled external perturbations. A collapse of the measurements
done at different imposed stress $\sigma_0$ can be obtained by
plotting $\eta^* = \eta/\sqrt{1 - \frac{\sigma_0^2}{\sigma_D^2}}$ as a
function of the modulation stress rate $R_\sigma$ (insert of
Fig.~\ref{Gp}), in agreement with eq.~\eqref{eq:visc}.\\
%
--{\it Summary and discussion.} In this letter, we propose to consider
a new fluidization pathway that could apply to a large class of soft
glassy materials arrested dynamically in the solid phase. The
mechanism requires two generic features, memory effects and non-linear
flow-induced rejuvenation. Under external shear stress and below the
yield stress, small fluctuations around the mean shear, accumulate
tiny irreversible strains over a long time and lead to \emph{secular}
drifts~\cite{Hinch1991} that can be viewed as an effective viscous
response. Even though the derivation was explicitly done on a simple
macroscopic rheological model, the existence of a secular term
yielding a finite material fluidity, is a generic feature coming out
from any model mixing ageing and non-linear rejuvenation
process~\cite{Voigtmann2014}. The underlying mechanism at work is in
principle very different from a thermal activation or any equivalent
mechanism accounting for stress fluctuations as an effective
temperature~\cite{Berthier2000,Marchal2009}. In the last case, the
amplitude of the fluctuations must help to overcome a barrier or a
threshold. In our case, fluidization stems from a dynamical
bifurcation of the rheological equations as a very general feature of
a dynamical system hosting processes working at very different time
scales. It would be interesting to see how in more sophisticated SGR
models with memory kernels accounting for ageing \cite{Fielding2000},
the equation's dynamics solved for similar driving conditions, would
also give a secular drift.
Evidence and quantitative assessment of the effect was brought for a
granular packing submitted to controlled stress modulations below the
Coulomb threshold. We related quantitatively the effective viscosity
to the inverse of the stress modulation rate and have shown that the
viscosity decreases significantly when approaching the dynamical
threshold.  Note that in spite of resemblances, this phenomenon is a
priori different from another fluidization process occurring when a
granular packing is placed in contact with a fluidized shear
band~\cite{Nichol2010,Reddy2011}. In the last case, theoretical
analysis and numerical simulations show that the induced creeping
process comes from a non-local stress relaxation, from the flowing
part to the material
bulk~\cite{PF09,Kamrin2012,Bouzid2013,Henann2014}. The generality of
the scenario, mixing two generic features of glassy system make it
suitable to be tested experimentally on many other practical situation
like colloidal glasses, pastes, clays or even glass-former molecular
systems, which actually may turn out to be of practical importance to
assess the stability and reliability of structures strained externally
in their environment over very long time scales. Finally, an important
question remains on the plastic relaxation modes involved in the
material strain in the context of this scenario (localized or extented
?). For granular matter this is the scope of a future report
\cite{Pons2015}. \\
--{\it Acknowledgments.} EC, AP and TD acknowledge the ANR grant
"Jamvibe-2010" and a CNES-DAR grant; AP, a CNES Post-doctoral
financial support; AA and JC, ``Action incitative'' funding from UR1.
\section*{APPENDIX: Numerical simulations}
\subsection*{Models comparison}
\begin{figure}[h]
\includegraphics[width=0.7\linewidth]{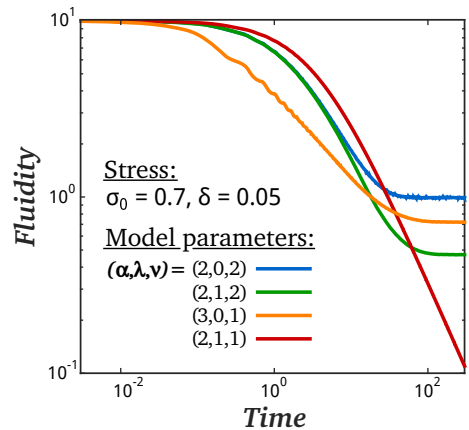}
\caption{Fluidity as a function of time obtained from the numerical integration of Eq.~\ref{eq.Ap4} for different set of exponents and subjected to a sinusoidal perturbation ($\sigma_0=0.7, \delta = 0.05$).}
\label{fig.Ap1}
\end{figure}

In Ref.~\cite{Derec2001}, a general form for the equation governing
the fluidity is proposed, coming from a ``Landau-type'' expansion:
\begin{equation}
\frac{\partial f}{\partial t} = -a \left( 1 - \left(
\frac{|\sigma|f}{|\dot{\gamma}|} \right)^{\lambda}
\frac{|\dot{\gamma}|^{\nu - \epsilon}}{f^{\nu}} \right) f^{\alpha}\label{eq.Ap4}
\end{equation}
where the higher orders of $f$ in the expansion have being dropped
because we work in the pasty phase where $f$ is small.  We also only
study the cases when $\epsilon = 0$ because we want to study a yield
stress fluid (see~\cite{Derec2001}).  The analytical study presented
in our letter treats the case $(\alpha, \lambda, \nu,\epsilon) =
(2,0,2,0)$.  Nevertheless, the underlying mechanism which leads to a
sub-threshold rejuvenation of the fluidity originate from the
$|\dot{\gamma}|^{\nu - \lambda}$ term ($\epsilon = 0$), so that when
$\nu \neq \lambda$, the subthreshold fluidization should always be
observed.  We demonstrate this using numerical integrations for
different sets of exponents. Fig.~\ref{fig.Ap1} shows the evolution in
time of $f$ for four set of $(\alpha, \lambda, \nu)$ keeping
$\epsilon=0$. We set $a=1$ and impose a sinusoidal stress
($\sigma_0=0.7, \delta = 0.05$).

We obtain a finite fluidity whenever $\nu \neq \lambda$. On the
contrary, when $\nu = \lambda$ (red curve of Fig.~\ref{fig.Ap1}, case
$(\alpha, \lambda, \nu) = (2,1,1)$), the creep remains logarithmic in
presence of small perturbations because the equation becomes:
\begin{equation*}
\frac{\partial f}{\partial t} = -a \left( 1 - |\sigma|\right)
f^{\alpha}
\end{equation*}
and we have always $|\sigma(t)|\ll1$ as the perturbation is well below
the threshold. Consequently no fluidisation can be observe as the
perturbation is not strong enough to pull the system over the
threshold.

Varying $a$ or $\sigma$ does not affect the general behavior of the
system as long as $\sigma(t)$ remains below 1.

\subsection*{Response to a random forcing}

\begin{figure}[h]
\begin{center}
\includegraphics[width=.7\linewidth]{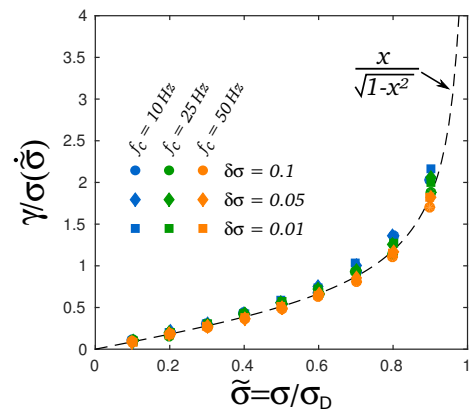}
\caption{Results of numerical simulation using noise instead of a
  regular oscillation as a perturbation: Strain rate in the steady
  state divided by the standard deviation of the stress rate ($\sigma(\dot{\tilde{\sigma}})$) as a function of the imposed mean
  stress normalized by $\sigma_D$. Results for $a=0.1$ and an initial fluidity of $f_0=10s^{-1}$.}
\label{fig.Ap2}
\end{center}
\end{figure}

We also test numerically the response of the model presented in the
main text to a stress modulated by random fluctuation.  We found that
such modulation has the same overall effect than a regular
perturbation.  Fig.~\ref{fig.Ap2} shows the results of the numerical
integration of eq.~\ref{EquFluStress} and~\ref{eqfluidity2} using
$\sigma(t) = \sigma_0 + \xi(t)$ where $\xi(t)$ is a noise presenting a
uniform frequency distribution between 0 and $f_c$ and whose standard
deviation equals $\delta \sigma$. A sub-threshold fluidization is
recovered for all the set of parameters we have tested.  By
normalizing the strain rate obtained in the steady-state by the
standard deviation of the stress rate
($\sigma(\dot{\tilde{\sigma}})$), we obtain a collapse of the data for
$\sigma_0 \ll \sigma_D$. One can note that the collapse perfectly on
our analytical solution
\begin{equation*}
\dot{\gamma}_{ \infty} = \frac{ \dot{\Sigma}
  \frac{\sigma_0}{\sigma_D}}{\sqrt{1 - \frac{\sigma_0^2}{\sigma_D^2}}}
\end{equation*}
in which $\dot{\Sigma}$ is a characteristic stress
rate. $\dot{\Sigma}$ corresponds to $\omega\delta/\sqrt{2}$ and the
standard deviation of the stress rate for sinusoidal modulations and
random modulations, respectively.


\end{document}